\documentclass[12pt]{iopart}
\usepackage{a4wide}

\usepackage{amssymb}
\usepackage[T1]{fontenc}
\usepackage{color}
\usepackage[utf8]{inputenc}

\usepackage{hyperref}

\begin{document}

\title{How can the nucleus be lighter than its constituents?}

\author{N.-E.\ Bomark} 

\address{Department for Natural Sciences, University of Agder, Universitetsveien 25, 4630 Kristiansand, Norway}

\ead{nilserik.bomark@gmail.com}

\begin{abstract}

The fact that the nucleus is lighter than its constituents, seems rather strange. How can the whole have a smaller mass than its 
components?

To get some intuition about how this is possible, one can look at a simpler more familiar system exhibiting the same phenomena; the hydrogen atom. It turns out that the same is true here; the hydrogen atom is a little bit lighter than the sum of its constituents. This difference corresponds to the ionisation energy of hydrogen.

This observation allows a simple explanation for how this is possible; the destructive interference between the electric fields of the proton and electron causes a reduction in the energy of the electric field and hence a reduction in the contribution to the mass of the system from the electric field energy.

The same explanation can be extended to the nucleus, thus giving some intuition about how this mass reduction happens. In the process, the idea of classical particles being ball-like objects with localised properties, is challenged.

\end{abstract}


\submitto{\EJP}

\maketitle

\section{Introduction}

It is a well known fact that the mass of an atomic nucleus is smaller than the sum of the masses of the protons and neutrons it is made of. This is 
something that we learn in the introduction to nuclear physics, but it is rather strange. How can the whole have a smaller mass than its 
components? It is a bit like taking a car and cutting it into pieces and the pieces would weigh more than the car you started with.

In this paper, we examine how this can be better understood by looking at the equivalent effect in a hydrogen atom. It will be demonstrated that the destructive interference in the electric fields from the proton and electron can explain the mass reduction and this gives us the intuition required to also better understand what is happening in the nucleus. It should be noted that this is not a new explanation; we do know that the mass discrepancy is the same as the binding energy of the system. However, seeing that this binding energy can be viewed as the energy lost due to the destructive interference between the electric fields, may provide a more intuitive picture of where the mass disappears.

After the link between the mass reduction in the nucleus and the equivalent effect in chemical bonds, is established in section~\ref{sec:chem}, we look in some detail at the hydrogen atom in section~\ref{sec:hydrogen}. We then in section~\ref{sec:Emass}, explain how this is due to the reduction of the energy in the electric fields and in section~\ref{sec:nucleus} we apply the same logic to the nucleus. Finally we discuss some implications of this in section~\ref{sec:Impl} before we conclude.

\section{The mass of a nucleus}

The source of the energy in radioactive decays was for a long time a mystery. Now we know that this comes from the atomic nucleus having a 
smaller mass after the decay than before. Equivalently we can regard the energy as the binding energy of the nucleus.

From any table of nuclide masses, we can see that the mass of a nucleus is always smaller than the sum of the masses of the protons and 
neutrons and the difference is what we call the binding energy. This raises a question, how can the nucleus weigh less than its constituents?

This question is usually not discussed, perhaps some mention of the strong nuclear force and maybe some discussion of the 
semi-empirical mass formula from the MIT bag model is given (see for instance~\cite{UP}), but other than that we do not get any intuitive understanding for how this is possible.

Coming from classical physics, this is very counter-intuitive. If 10 balls, each with a mass of 100 g are put in a basket, and we discover that 
the basket with all the balls has a mass of 950 g, we would just assume one ball fell out.

\section{Energy conservation and chemistry}\label{sec:chem}

Before we attempt to explain this we should examine a bit closer what we do know. First we shall look at the relation between energy and mass 
as this is at the heart of our discussion.

In special relativity, the energy, $E$, the mass, $m$, and the momentum, $\vec p$, are related as $E^2=(mc^2)^2+(\vec p c)^2$. If we set the momentum to zero we get the famous $E=mc^2$ and this tells us that for anything with no momentum, energy and mass are equivalent concepts. Somewhat more catchy; mass is energy that does not move.

Let us look at some reaction, nuclear or chemical, that releases some energy. Before the reaction all energy takes the form of mass; we assume that all movements are negligible, then mass and energy are the same thing. Afterwards we again have mass, but also the released energy. This released energy then must come from a reduction in the overall mass. There is no other way for energy to be conserved.

This line of thought tells us that the energy released in any nuclear decay must come from the mass of the mother nucleus and as an extension of, the mass of the nucleus must be smaller than its constituents, otherwise it would be unstable.

There is one other consequence; the same must be true for chemical reactions, any exothermic reaction must get its energy from the mass of the initial reactants. Since the reaction energies in chemical reactions are much lower, the change in mass is typically not measurable, so chemists can keep teaching mass conservation, but for our purposes the tiny deviation from this is important. As we will see, this gives us an opening for a better understanding of where the mass disappeared.

Before we dive into that, let us note that this also must happen in gravitational systems. If we look at the Earth-Moon system, the combined system must have a total mass that is lower than the combined mass of the celestial bodies by an amount equal to the gravitational binding energy. This binding energy is given by the absolute value of the gravitational potential energy, $U_G= -GM_{\rm Earth}M_{\rm Moon}/{R}$, where $R=3.8\times 10^{8}$ m, is the Earth-Moon distance.
This gives us a binding energy which, if divided by $c^2$, equals $8\times 10^{11}$ kg. In other words, the combined system is 800 million tons lighter than the two bodies would have been if separated. Though 800 million tons may sound like a big mass, it is not; we must compare it to the mass of the Earth which is $6\times 10^{24}$ kg, so this mass reduction really is too small to have measurable consequences.

\section{The hydrogen atom}\label{sec:hydrogen}
The mass change in chemical reactions are typically not measurable, but there are two exceptions, the hydrogen and helium atoms. In these two cases the masses involved are measured with phenomenal precision.

Let us first look at the hydrogen atom. If we compare the mass of a free proton and a free electron with the mass of a hydrogen atom, we get a mass difference~\cite{pdg, Atomer},
\begin{equation}
	\Delta M = M_p+M_e-M_H = 1.43\times 10^{-8} {\rm u}, 
\end{equation}
which, if converted to energy through $E=mc^2$, becomes 13.3 eV. This is, within the experimental uncertainty, consistent with the 13.6 eV known to be the ionization energy of hydrogen.

If we do the same thing for a helium atom, calculate the mass difference~\cite{NIST},
\begin{equation}
	\Delta M = M_\alpha+2M_e-M_{He} = 8.48\times 10^{-8} {\rm u}
\end{equation}
and convert to energy, we get 79 eV which is the measured double ionization energy or binding energy for helium~\cite{NISTHe}.

These are clear examples that this mass reduction in a combined system is not unique to the nucleus and we can not hide behind the complexities of the strong nuclear force. We should be able to understand this with just electromagnetism; after all, the atom is held together entirely by electric forces. It is therefore time to look closer at the electric field.

\section{The mass in the electric field}\label{sec:Emass}

An electron is always accompanied by an electric field, that is what it means to be electrically charged. Since this electric field is inseparable from the electron, we must consider it an integral part of the electron and expect it to have an impact on all of the electrons properties.

It is a known fact that electric fields have energy, and specifically an electric field, $E$, possesses an energy density, $u_E$, given by $u_E=\epsilon_0 E^2 /2$.

The electric field around an electron therefore comes with an energy, and since that energy does not move it must appear as a mass. This mass is associated with the electron through its electric field and must therefore contribute to the total mass of the electron.

This is an important point, the mass of an electron is not entirely localised but partly distributed throughout the electric field surrounding it. It now looks more feasible to explain the disappearance of mass when an electron and a proton are combined into hydrogen, if some of the mass is spread out in the electric field, we need to look closer at the electric field of the atom.

When the proton and electron are combined, their electric fields outside the atom interfere destructively cancelling each other out, this is why the atom is electrically neutral. This cancellation also removes some of the energy of the electric fields and thus reducing the mass of the combined system. To be more precise, let us estimate how large the reduction is.

From Coulombs law we have the electron field, $E$, around an electron as $E= {e_c}/{(4\pi \epsilon_0 r^2)}$. As already mentioned, the 
energy density of an electric field is $u_E=\epsilon_0 E^2 /2$. To get the total energy, $U_E$, we need to integrate this, if we integrate 
from a small distance $r_0$ from the electron\footnote{There is a divergence for small distances so we must exclude the region closest to 
the electron. We also know that there are quantum corrections to the electric field for $r \lesssim 10^{-13}$ m so we cannot trust our 
calculations there. This is an interesting topic in itself and can elucidate the problem of infinities in quantum field theories, but that is 
a discussion for another time, here we are interested in longer distances.} to infinity we get,
\begin{equation}\label{eq:energy}
 U_E(r_0) = \frac{e_c^2}{8\pi \epsilon_0}\frac{1}{r_0}.
\end{equation} 

If we now assume that the cancellation takes place from a distance of $10^{-10}$ m and outwards\footnote{What happens inside the atom is hard to see, but from numerical estimates it seems we have about as much constructive as destructive interference between the fields and hence no effect on the overall mass.} 
(approximate size of a hydrogen atom), we get  $U_E(r_0) \approx 7$ eV. Since we are looking at the electric fields more than $10^{-10}$ m away from the particles, we can ignore the inner structure of the nucleus and the fields from the electron and proton are identical in strength and opposite in direction (this is why they cancel each other). This means that the energy removed from the electric field of the proton is identical to the energy removed from the electron's electric field and hence the total reduction in rest energy (or mass) is twice the above result, so around 14 eV, in surprisingly good agreement with the 
measured mass reduction.

We can also use this logic on two particles with same sign charge, then the interference will be mostly constructive, increasing the mass of the combined system. This is an alternative explanation for why same charges repel; since the combined system gets higher mass than the particles, it will spontaneously break apart into separated particles. Though there is no new physics here, just a reformulation of what we know, it can be beneficial for students to see different accounts of the same phenomena and this does explain how Coulomb repulsion increases the mass of heavier nuclei and eventually makes them unstable.

\section{The mass of the nucleus revisited}\label{sec:nucleus}

Now that we can explain the mass reduction in the hydrogen atom it is time to come back to the nucleus. Although the nucleus is a much more 
complicated system, we can imagine that the mass reduction is due to interference in the fields holding the nucleus together, similarly to 
the case of the hydrogen atom. One could leave it with this analogy, saying that what happens in the nucleus is some more 
complicated version of what we have seen in the hydrogen atom, but let us see if we can do better.

To do this we need a model for how the nucleus is held together. Let us use the simplest model due to Yukawa~\cite{Yukawa:1935xg} where we have protons and 
neutrons held together by a spin-0 force field. In modern terms we can view this field as a collection of meson fields (as a matter of fact, 
it is from Yukawa we have the word meson). These fields have mass and therefore very short range, but that is not so important for us. What 
is important though is that the most important of them, in contrast to the spin-1 electromagnetic field, are spin-0 fields. It turns out that this means the force is attractive~\cite{Zee}, which manifests in a negative energy density. 

With a negative energy density, the constructive interference we will get from the fields around the nucleons will create a more negative 
contribution to the overall mass and that is how the mass is reduced in the combined system.

\section{The nature of elementary particles}\label{sec:Impl}

In addition to the more intuitive understanding of the disappearance of mass in combined systems, this explanation can help shedding light on some more elusive properties of elementary particles. 

When talking about elementary particles such as electrons and protons, it is easy to fall for the idea of these particles being small ball-like objects where all the properties are localised in this little ball, after all this is how they are usually depicted. This becomes problematic when confronted with quantum mechanics.

Realising that not even classical particles can be described by such balls --- their properties, such as mass, are not localised, but at least partly distributed throughout the electromagnetic (and other) fields around them --- might make it easier to accept that this is even less possible in quantum mechanics.

From a more advanced point of view, this also illustrates that an electron cannot be thought of separately from its electric field, the electric field is an integral part of the electron and affects its properties. This comes back in quantum field theory when interactions between fields renormalise the properties of those fields through loop diagrams.

\section{Conclusions}

It should now be clear that the lightness of the nucleus in comparison with its constituents, is not as mysterious as one might think when first encountering it. The existence of direct analogs in the hydrogen and helium atoms (and all chemical systems, though we cannot measure that) demonstrates that this is not some  strange phenomena unique to nuclear physics, but a general property of all bound systems of any kind.

The fact that the mass reduction can be explained in simple terms through the destructive interference between electric fields, should make this more comprehensible and easier to accept.

Moreover, the realisation that the source of the mass reduction is in the mass contribution from the electric field energy, shows that we must acknowledge the electric field as an integral part of the particles and thus challenges the intuitive picture of elementary particles as small bouncing balls. This could prove useful at a later stage when quantum physics is introduced and the idea of ball-like particles become untenable.

\vspace*{0.2 cm}
%{\bf Acknowledgements.}

\section*{References}
%%%%%%%%%%%%%%%%%%%%%%%%%%%%%%%%%%%%%%%%%%%%%%%%%%%%%%%%%%%%%%%%%%%%%%

\end{document}